\documentstyle[prb,preprint,aps,12pt]{revtex}
\tightenlines
\title{Dissimilar Ferromagnetic Layer Dependence of Hot Electron Magnetotransport}
\author{Jisang Hong}
\address{Max-Planck-Institut f{\"u}r Mikrostrukturphysik, Weinberg 2, D-06120 Halle, Germany}
\begin{document}
\maketitle
\begin{abstract}
The material dependence of hot electron  magnetotransport in a spin-valve transistor has been theoretically explored. We calculate the parallel and anti-parallel  collector current changing the types and relative spin orientation of the ferromagnetic layers. The magnetocurrent has been presented as well. In this calculations, spin dependent self energy effect of hot electron in ferromagnetic materials has been taken into account. The results show that the magnetotransport property strongly depends on the combination of different ferromagnetic metal layers since the hot electron has different inelastic scattering strength in each material, and the hot electron spin polarization enters importantly into the magnetocurrent at finite temperatures. This calculations may suggest the guide for searching the best structural combination of hot electron magnetoelectronic device such as a spin-valve transistor. 
\end{abstract}
\pacs{ 73.30.Ds,75.30.-m,75.25.+z, 85.70.-w}
\newpage
\newcommand{\eb}{\begin{eqnarray}}
\newcommand{\ee}{\end{eqnarray}}
\section{Introduction}
Discovery of giant magneto resistance (GMR) \cite{GMR} in magnetic multilayer structure has brought great interests in the study of magnetic thin film structure because of fundamental interests as well as technological importance. For example, magnetic tunneling junction (MTJ) \cite{junction} is being widely explored for the purpose of magnetoelectronic device. Interestingly, a new type of promising magnetoelectronic device, so called a spin-valve transistor (SVT) \cite{valve}, has been suggested by Monsma {\it et al} as well. Unlike the conventional magnetic tunneling junction, one encounters different structure \cite{structure} and properties in this spin-valve transistor. For instance, a spin-valve transistor has typically $Si/N_1/F_1/N_2/F_2/N_3/Si$ structure where $N_i (i=1..3)$ represents normal metal, and $F_i (i=1,2)$  stands for ferromagnetic layer. In this structure electrons across the Schottky barrier (emitter side) penetrate the spin-valve base, and the energy of hot electrons is above the Fermi level of metallic base. Thus, {\it hot} electron magnetotransport should be taken into account when one explores the spin-valve transistor.  

Transport property of hot electrons is different from that of Fermi electrons. In the magnetic tunneling junction electrons near the Fermi level mostly contribute to the tunneling current, and spin polarization of these Fermi electrons strongly depends on the density of states near the Fermi level. In contrast, the {\it hot} electron transport is related to the density of unoccupied states above the Fermi level, and it has an exponential dependence on electron inelastic mean free path \cite{inelastic}. The exponential dependence on the inelastic mean free path results in many interesting features in hot electron device such as the spin-valve transistor incorporated with magnetism. For instance, a collector current has strong sensitive to the relative spin orientation in the ferromagnetic layers because of its exponential dependence on the inelastic mean free path , and magnetocurrent does not depend on any spin independent attenuation. These properties indicate that the spin-valve transistor can be a very favorable candidate for magnetoelectronic device. Very recently, Jansen {\it et al} \cite{Jansen} reported temperature dependence of collector current and magnetocurrent changing the relative spin orientation in ferromagnetic layers. They obtained huge magnetocurrent even at room temperature and unusual temperature dependence of collector current. A spin mixing mechanism due to thermal spin waves is suggested to account for their observation beyond 200 K. Regarding the issue raised in their paper, theoretical calculation \cite{theoretical} have been presented to explore the relative importance of spin mixing and hot electron spin polarization to the collector current at finite temperatures. Interestingly, the theoretical calculations suggest that the hot electron spin polarization has a substantial contribution to the hot electron magnetotransport in the spin-valve transistor. 

As remarked earlier, the spin-valve transistor has been suggested as another type of magnetoelectronic device for real applications. A major advantage of this structure is in the sensitivity of collector current to the relative spin orientation of the ferromagnetic layers because of an exponential dependence of the collector current on the inelastic mean free path. On the other hand. it has a serious difficulty for real applications since the output collector current is too small to the present time. According to the measurement \cite{Jansen}, when the relative spin orientation of the ferromagnetic layers is parallel the collector current (parallel collector current) is roughly 8 - 10 nA, and the anti-parallel collector current is around 2 - 3 nA with 2 mA input current. Thus, one of major efforts in this area is to find the best structure for larger output collector current. Bearing this in mind, we shall study the dissimilar ferromagnetic layer dependence of the hot electron magnetotransport at finite temperatures, and search the best structural combinations of ferromagnetic materials in the spin-valve transistor.

\section{Model}
We shall explore the magnitude of output collector current as well as the magnetocurrent changing the combination of the ferromagnetic layers. In this model calculations the normal metal layers are considered to be the same material, hence we will focus our interests on the magnetotransport in the ferromagnetic layers. We assume that the Schottky barrier has no spatial distribution. Including a spatial distribution of Schottky barrier \cite{barrier} of course will have an influence on the magnitude of output collector current. However, since the  Schottky barrier exists at the interface of the semiconductor and normal metal layer it does not have any spin dependence. Then to treat the Schottky barrier as a very ideal case dose not change essential physics for our purpose in this work. The energy of the injected electrons across the Schottky barrier into the spin-valve base is around 0.9 eV above the Fermi level \cite{Jansen}, we therefore take the energy of the hot electrons in this model calculations as 1 eV above the Fermi level. Now, the the central issue of this work is to analyze the magnitude of the collector current and the magnetocurrent at finite temperatures depending on the dissimilar ferromagnetic layer combination. For the sake of argument we denote the spin-valve transistor structure as $F_1/F_2$ because normal metal layers are not changed at all in this model calculations. The Fe/Fe, Ni/Fe, and Ni/Ni structures will be explored in this work because the Fe has the largest magnetic moment, and the Ni has the smallest one in 3d ferromagnetic transition metals. 

Once the hot electrons start to penetrate the spin-valve base we then need to explore the Green's function $G_\sigma({\vec{k},E})$, which describes the propagation of the electron (spin up and spin down) in each material. We can write this as
\eb
G_\sigma({\vec{k},E})=\frac{1}{E-\epsilon_\sigma(\vec{k})-\Sigma_\sigma({\vec{k},E})}
\ee
The theoretical calculations of spin dependent self energy \cite {path} including the effect of spin wave excitations, Stoner excitations, and various spin non-flip precesses in the ferromagnets have been presented. The theoretical calculations show that the self energy $\Sigma_\sigma({\vec{k},E})$ has a strong spin dependence in ferromagnets, so that the inelastic mean free path is spin dependent in the ferromagnetic materials. We define $\gamma_{M_i}(T)$ to describe the spin dependent inelastic scattering effect of majority spin electrons in ferromagnetic material $F_i$ at finite temperatures and $\gamma_{m_i}(T)$ for minority spin electrons. We can write this as $\gamma_{M_i(m_i)}(T)=exp[-w_i/l_{M_i(m_i)}(T)]$ where $l_{M_i(m_i)}(T)$ is the inelastic mean free path of majority (minority) spin electron in ferromagnetic layer $F_i$ at temperature T, and $w_i$ is the thickness of that material. One can also relate these $\gamma_{M_i}(T)$ and $\gamma_{m_i}(T)$ to the hot electron spin polarization. In this work we will adopt the results presented in Ref \cite{path}. Definitely, there will be an attenuation when the hot electrons are passing through the normal metal layer $N$ as well as ferromagnetic layer $F_i$. We denote the attenuation in the normal metal layer $N$ as $\Gamma_N(T)$. As remarked in the above, the current has an exponential dependence on the electron inelastic mean free path, therefore the inelastic scattering effect in normal metal layer has the same influence on any combination of ferromagnetic layers giving the exactly same contribution to the parallel and anti-parallel collector current. This exponential dependence of the collector current on the inelastic mean free path enables us to focus our interests only on the ferromagnetic layers.

Since the hot electrons are not spin polarized until they reach the first ferromagnetic layer, we therefore can say that $N_0/2$ spin up and spin down electrons injected into the spin-valve base per unit time per unit area, respectively. After they enter into the ferromagnetic layer the hot electrons will suffer from the strong spin dependent inelastic scatterings. Then, $N_0/2\gamma_{M_i(m_i)}(T)$ electrons penetrate the first ferromagnetic layer if they are majority (minority) electrons. One can describe the attenuation in the second ferromagnetic layer in the same way. It should be noted again that recent theoretical calculations \cite{theoretical} suggest that the hot electron spin polarization to the collector current has a substantial contribution to the collector current compared to that from spin mixing effect due to thermal spin waves at finite temperatures . In this paper, thus we will consider the effect of hot electron spin polarization at finite temperatures. Keeping all these in mind, one can write the collector current in parallel configuration at finite temperatures
 
\eb
\tilde{I}_c^P(T)=\frac{N_0}{2}\Gamma^3_N(T)\gamma_{M_1}(T)\gamma_{M_2}(T)[1+\frac{\gamma_{m_1}(T)}{\gamma_{M_1}(T)}\frac{\gamma_{m_2}(T)}{\gamma_{M_2}(T)}]. 
\ee
Similarly, in the case of anti-parallel
\eb
\tilde{I}_c^{AP}(T)=\frac{N_0}{2}\Gamma^3_N(T)\gamma_{M_1}(T)\gamma_{M_2}(T)[\frac{\gamma_{m_1}(T)}{\gamma_{M_1}(T)}+\frac{\gamma_{m_2}(T)}{\gamma_{M_2}(T)}].
\ee
As remarked in the above, $\gamma_{M_i}(T)$ and $\gamma_{m_i}(T)$ can be related to the hot electron spin polarization $P_{H_i}(T)$ in ferromagnetic layer $F_i$ at finite temperatures. One can write this
\eb
\frac{\gamma_{m_i}(T)}{\gamma_{M_i}(T)}=\frac{1-P_{H_i}(T)}{1+P_{H_i}(T)}.
\ee 
From this expression, most generally the $\gamma_{M_i}(T)$ and $\gamma_{m_i}(T)$ can be expressed as
\eb
\gamma_{M_i}(T)=g_i(T)(1+P_{H_i}(T))
\ee
and
\eb
\gamma_{m_i}(T)=g_i(T)(1-P_{H_i}(T))
\ee
where $g_i(T)$ is a function of temperature T, and this function $g_i(T)$ enters into the $\gamma_{M_i}(T)$ and $\gamma_{m_i}(T)$ simultaneously. With these relations, we can obtain the expression of the collector current at finite temperatures 
\eb
\tilde{I}_c^P(T)&=&\frac{N_0}{2}\Gamma^3_N(T)g_1(T)g_2(T)(1+P_{H_1}(T))(1+P_{H_2}(T)) \nonumber \\
&& \times [1+\frac{1-P_{H_1}(T)}{1+P_{H_1}(T)}\frac{1-P_{H_2}(T)}{1+P_{H_2}(T)}]
\ee
and
\eb
\tilde{I}_c^{AP}(T)&=&\frac{N_0}{2}\Gamma^3_N(T)g_1(T)g_2(T)(1+P_{H_1}(T))(1+P_{H_2}(T)) \nonumber \\
&& \times [\frac{1-P_{H_1}(T)}{1+P_{H_1}(T)}+\frac{1-P_{H_2}(T)}{1+P_{H_2}(T)}]
\ee
With these quantities, one can also readily calculate magnetocurrent 
\eb
MC(T)=\frac{\tilde{I}_c^P(T)-\tilde{I}_c^{AP}(T)}{\tilde{I}_c^{AP}(T)}
\ee

The central issue of this work is to understand the magnitude of the collector current and the magnetocurrent depending on the dissimilar ferromagnetic layer combination. For quantitative analysis, it is necessary to know the temperature dependence of inelastic mean free path in the ferromagnetic layers as well as in the normal metals. Here, it is of importance to note that the attenuation of low energy electron in the normal metal is around $100 \AA$ \cite{normal}. It is several times greater than that in the ferromagnets \cite{path}. This implies that the inelastic scattering in the ferromagnetic layers enters importantly into the magnetotransport. Hence, in this model calculations it is assumed that the $\Gamma_N(T)$ is temperature independent. Now, one therefore needs to understand the quantity $g_i(T)$ in Eqs. (5) and (6). From the Eq. (6), the relation $exp[-w_i/\lambda_{m_i}(T)]=g_i(T)(1-P_{H_i}(T))$ is easily understood, and since physically it is clear that the inelastic mean free path will be decreasing with temperature T, the condition $exp[-w_i/\lambda_{m_i}(T)] \le exp[-w_i/\lambda_{m_i}(0)]$ should be satisfied. Thus, one can clearly understand that $g_i(T)$ is a decreasing function with temperature T with its maximum value $exp[-w_i/\lambda_{M_i}(0)] \times (1+P_{H_i}(0))^{-1}$ or $exp[-w_i/\lambda_{m_i}(0)] \times (1-P_{H_i}(0))^{-1}$. Although there is an example of lifetime measurement of Co \cite{Co} in the relevant energy ranges to the spin-valve transistor, it does not show any data about the temperature dependence. Since we have no reliable data about $g_i(T)$ at finite temperatures  $g_i(T)$ is replaced by $g_i(0)=exp[-w_i/\lambda_{M_i}(0)] \times (1+P_{H_i}(0))^{-1}$ to calculate the magnitude of the collector current. This condition implies that the calculated collector current in this model calculations will be the maximum magnitude of the collector current. Under this condition, the quantities  
\eb
I_c^P(T)&=&\frac{N_0}{2}\Gamma^3_N(0)g_1(0)g_2(0)(1+P_{H_1}(T))(1+P_{H_2}(T)) \nonumber \\
&& \times [1+\frac{1-P_{H_1}(T)}{1+P_{H_1}(T)}\frac{1-P_{H_2}(T)}{1+P_{H_2}(T)}],
\ee
\eb
I_c^{AP}(T)&=&\frac{N_0}{2}\Gamma^3_N(0)g_1(0)g_2(0)(1+P_{H_1}(T))(1+P_{H_2}(T)) \nonumber \\
&&[\frac{1-P_{H_1}(T)}{1+P_{H_1}(T)}+\frac{1-P_{H_2}(T)}{1+P_{H_2}(T)}]
\ee
will be explored in this work. One can easily understand that the magnetocurrent will be the same even with the above Eqs. (10) and (11) as in Eq. (9). 

\section{Results and discussions}
The inelastic mean free path in ferromagnetic materials is taken from the results of Ref. \cite{path}, for example, $\lambda_M(0)=27\AA$ for Ni and $\lambda_M(0)=19\AA$ for Fe at 1 eV above the Fermi level, and the inelastic mean free path in the normal metal layer is used as $90\AA$ in this model calculations. The thickness of the first ferromagnetic layer is taken as $60\AA$ and $30\AA$ for the second magnetic layer to have the different coercivity. Along with that, the thickness of the normal metal layer is taken as $35\AA$. We model the hot electron spin polarization at finite temperatures as $P_H(T)=P_0(1-[T/T_c]^2)$ and $P_H(T)=P_0(1-[T/T_c]^{3/2})$ where $P_0$ is the hot electron spin polarization at zero temperature, and $T_c$ is the critical temperature of ferromagnetic material. 

We now discuss the results of the model calculations. Fig. 1 displays both the parallel and anti-parallel collector current for different combinations with the $P_H(T)=P_0(1-[T/T_c]^2)$, and Fig. 2 represents the results with $P_H(T)=P_0(1-[T/T_c]^{3/2})$. Here, we have explored the Eqs. (10) and (11) dividing by $N_0$ because $N_0$ is a common factor both for parallel and anti-parallel collector current. One can clearly see that the parallel and anti-parallel collector current behave differently with temperature T in any combination. Since the $1+P_H(T)$ and $1-P_H(T)$ have the opposite property with temperature T, these two terms are competing each other and contributing differently to the collector current depending on the relative spin orientation of the ferromagnetic layers. Now, if we look at the magnitude of the collector current for different combinations we can see that the largest collector current can be obtained from the Ni/Ni, while the Fe/Fe structure produces the smallest collector current. For example, the anti-parallel collector current of Ni/Ni structure is almost seven times greater than that of the Fe/Fe over the whole temperature ranges. The parallel collector current also shows almost the same trends. In the discussion of the magnitude of the collector current we should note the inelastic mean free path in ferromagnetic layers. According to the spin dependent self energy calculations \cite{path}, the hot electron has very strong inelastic scattering strength in Fe than that in Ni. As a result, the inelastic mean free path in Fe is shorter than that in Ni. Thus, Ni/Ni combination produces the largest collector current. Fig. 3 represents the magnetocurrent with different temperature dependence of hot electron spin polarization. Ni/Ni combination displays the smallest magnetocurrent and the most rapid temperature variation while we obtain the largest output collector current. In contrast, the  Fe/Fe has the opposite property. This temperature dependence of magnetocurrent can be understood in terms of hot electron spin polarization at finite temperatures. Since the critical temperature of Fe is roughly twice higher than that of Ni, the hot electron spin polarization varies rapidly in Ni rather than in Fe. Therfore, this model calculations suggest that the spin dependent inelastic mean free path and hot electron spin polarization enter into the hot electron magnetotransport significantly. 

In conclusion, the magnitude of collector current and the magnetocurrent depending on the dissimilar ferromagnetic layer combination have been explored. We obtain that the Ni/Ni combination may produce the largest output collector current while the magnetocurrent shows the most rapid temperature variation displaying the smallest magnitude, and the Fe/Fe structure has the opposite property. By the virtue of the fact that the minority spin electrons suffer from the strong inelastic scattering due to spin waves, Stoner excitations, and various spin non-flip precesses in ferromagnets the hot electrons magnetotransport is affected substantially by the combination of dissimilar ferromagnetic materials and hot electron spin polarization as well. We hope that this work will stimulate further related issues such as the temperature dependence of hot electron inelastic mean free path in metals at low energy and hot spin polarization, and help to find the best structural combination in the spin-valve system.  
\section*{Acknowledgments}
I am most grateful to Dr. P.S. Anil Kumar for his comments and discussions during the course of this work. 

\newpage
\begin{figure}
\caption{The parallel and anti-parallel collector current in Eqs. (10) and (11) with $P_H(T)=P_0(1-[T/T_c]^2)$ dividing by $N_0$. The critical temerature $T_c$ has been taken as 1200 K for Fe and 630 K for Ni.}
\end{figure}
\begin{figure}
\caption{The parallel and anti-parallel collector current in Eqs. (10) and (11) with $P_H(T)=P_0(1-[T/T_c]^{3/2})$ dividing by $N_0$. The same critical temeratures have been used for Fe and Ni as in Fig 1.}
\end{figure}
\begin{figure}
\caption{The magnetocurrent at finite temperatures. The circle is the magnetocurrent with $P_H(T)=P_0(1-[T/T_c]^2)$, and the asterisk is for $P_H(T)=P_0(1-[T/T_c]^{3/2})$.}
\end{figure}
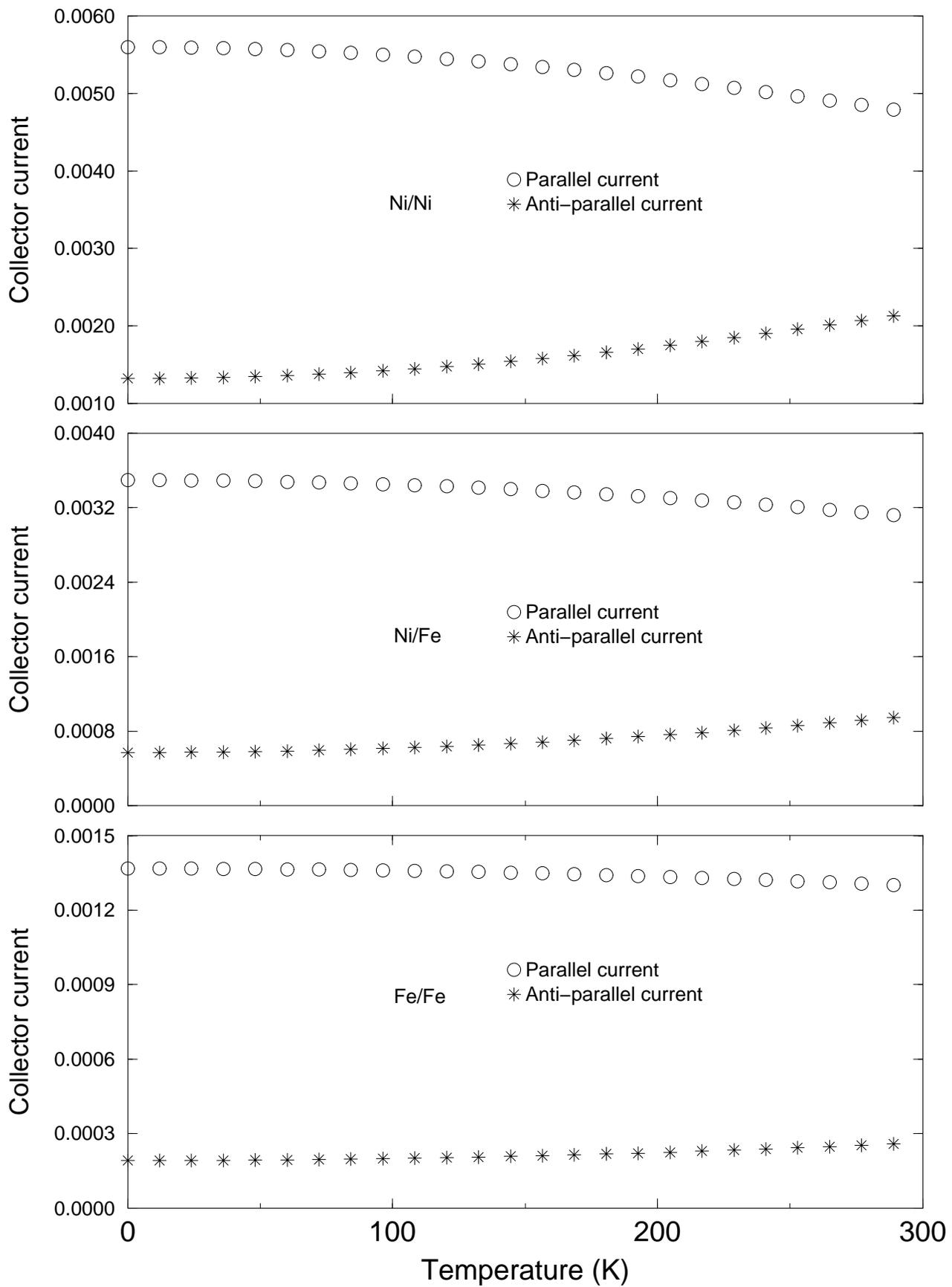
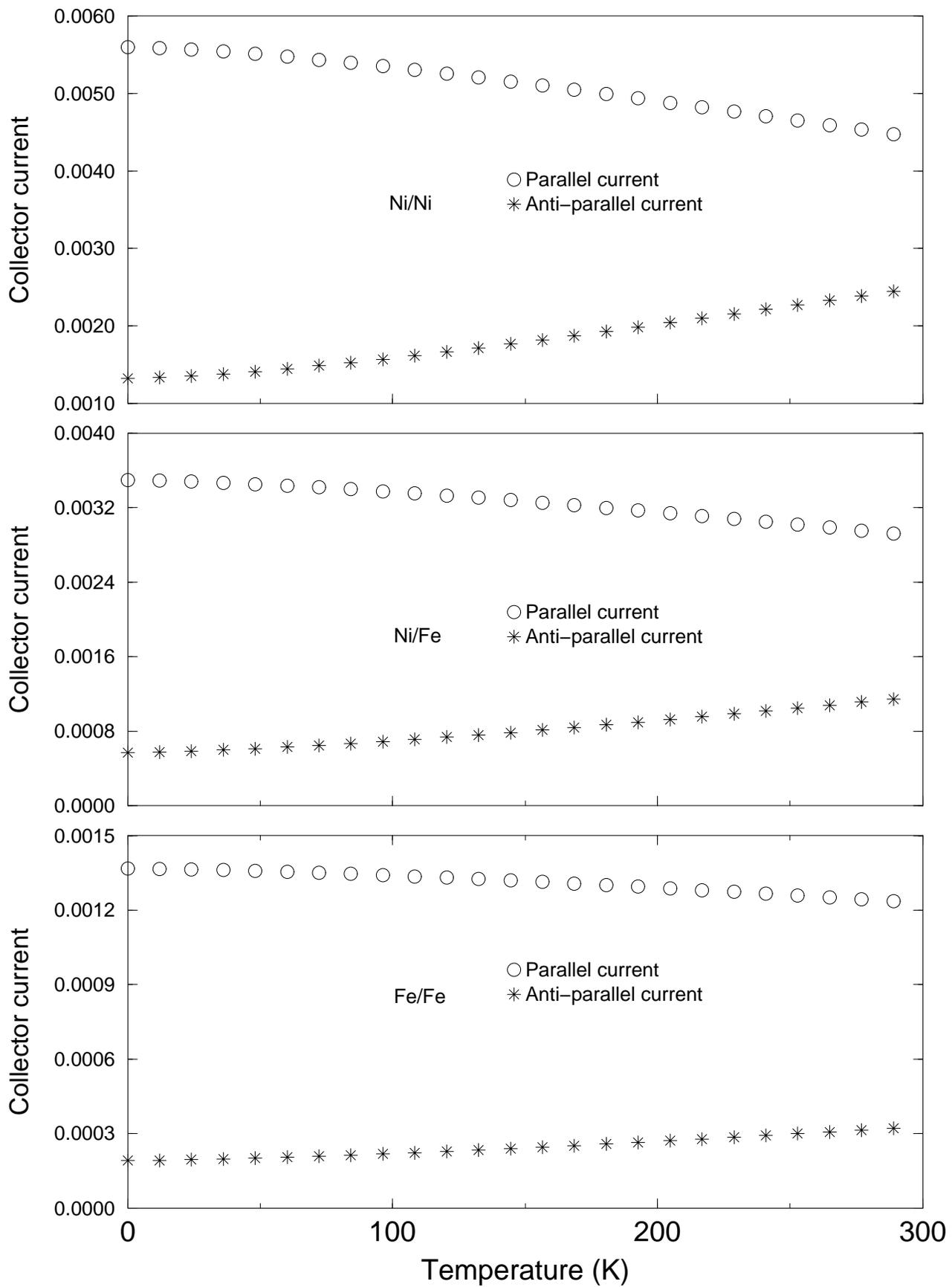
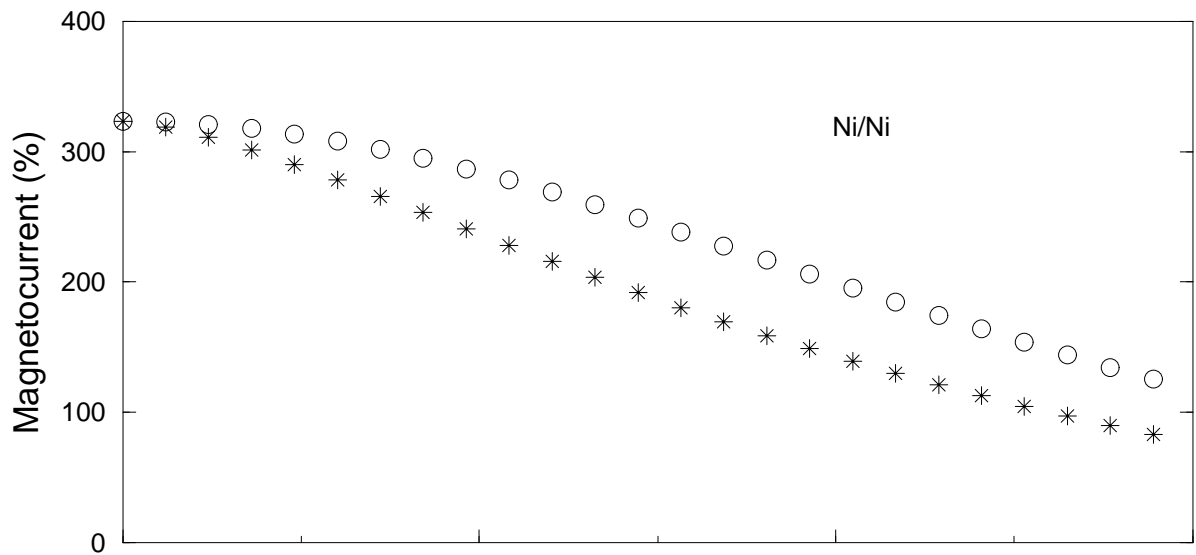
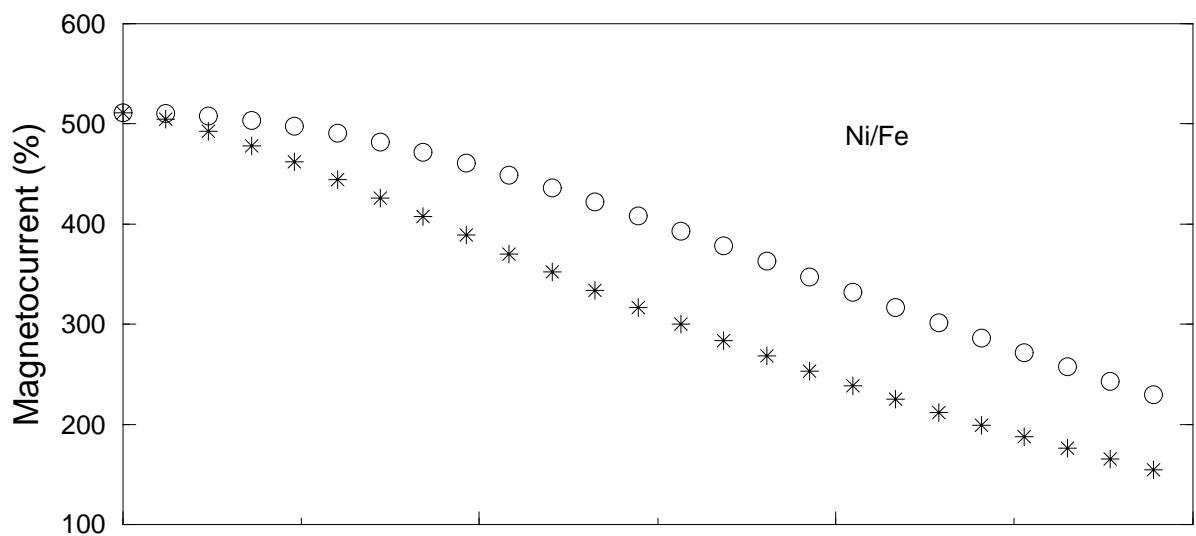
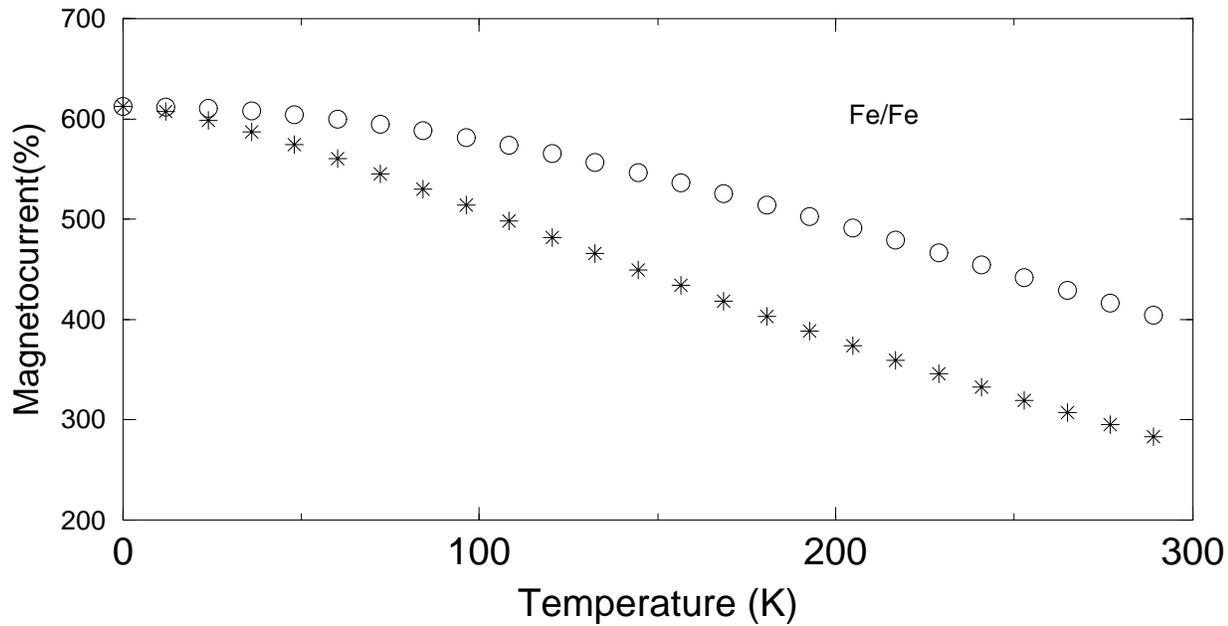
\end{document}